\documentclass[twoside,12pt]{article}
\usepackage{epsfig}
\def\Journal#1#2#3#4{{#1} {#2} (#4) #3 }

\newcommand{\be}{\begin{equation}}
\newcommand{\ee}{\end{equation}}
\newcommand{\bea}{\begin{eqnarray}}
\newcommand{\eea}{\end{eqnarray}}

\begin{document}
\title{ \vspace{1cm} Chiral symmetry and strangeness at SIS energies}
\author{Matthias F.M. Lutz,
\\
Gesellschaft f\"ur Schwerionenforschung GSI,
Postfach 110552\\ D-64220 Darmstadt, Germany}
\maketitle
\begin{abstract}
In this talk we review the consequences of the chiral SU(3)
symmetry for strangeness propagation in nuclear matter. Objects of crucial
importance are the meson-baryon scattering amplitudes obtained within
the chiral coupled-channel effective field theory. Results for
antikaon and hyperon-resonance spectral functions in cold nuclear matter are
presented and discussed. The importance of the $\Sigma (1385)$ resonance for
the subthreshold antikaon production in heavy-ion reaction at SIS is pointed out.
The in-medium properties of the latter together with an antikaon spectral function
based on chiral SU(3) dynamics suggest a significant enhancement of the
$\pi \,\Lambda \to \bar K \,N$ reaction in nuclear matter.
\end{abstract}



\section{Introduction}

The foundation of all theoretical attempts to describe and possibly predict hadron properties
in a dense nuclear environment should be Quantum Chromo Dynamics (QCD), the fundamental theory of
strong interactions. The QCD Lagrangian predicts all hadron interactions in terms
of quark and gluon degrees of freedom. At large momentum transfers QCD is extremely successful and predictive,
because observable quantities can be evaluated perturbatively in a straightforward manner.
At small momentum transfers, in contrast, QCD is less predictive so far due to its non-perturbative
character. It is still an outstanding problem to systematically unravel the properties of QCD in its low-energy phase
where the effective degrees of freedom are hadrons rather than quarks and gluons. That is a challenge since
many of the most exciting phenomena of QCD, like the zoo of meson and baryon resonances, manifest themselves
at low-energies. Also the fundamental question addressed in this talk 'how do hadrons change their properties
in dense nuclear matter' requires a solid understanding of QCD at small momentum transfer.

There are two promising paths along which one achieved and expects further significant progress. A direct evaluation
of many observable quantities is possible by large scale numerical simulations where QCD is put on a finite
lattice \cite{lattice:1}. Though many static properties, like hadron ground-state properties, have been successfully
reproduced by lattice calculations, the description of the wealth of hadronic scattering data is still outside the
scope of that approach \cite{lattice:1}. Also the determination of hadron properties at finite
and cold nuclear densities is still
not feasible on the lattice \cite{lattice:2,lattice:3}. Here a complementary approach, effective field theory, is more
promising at present. Rather than solving low-energy QCD  in terms of quark and gluon degrees of freedom,
inefficient at low energies, one aims at constructing an effective quantum field theory in terms of hadrons
directly. The idea is to constrain that theory by as many properties of QCD as possible and
establish a systematic approximation strategy. This leads to a significant
parameter reduction and a predictive power of the effective field theory approach.

An effective field theory, which meets the above criteria, is the so called Chiral Perturbation Theory ($\chi$PT)
applicable in the flavor $SU(2)$ sector of low-energy QCD \cite{book:Weinberg}. This effective field theory is based on a simple
observation, namely that QCD is chirally symmetric in the limiting case where the up and down current quark masses
vanish $m_{u,d} =0$. This implies in particular that the handedness of quarks is a conserved property in
that limit. There is mounting empirical evidence that the QCD ground state breaks the chiral $SU(2)$ symmetry
spontaneously in the limit $m_{u,d} =0$. For instance the observation that hadrons
do not occur in parity doublet states directly reflects that phenomenon. Also the smallness of the pion masses
with $m_\pi \simeq $ 140 MeV much smaller than the nucleon mass $m_N \simeq $ 940 MeV, naturally fits
into this picture once the pions are identified to be the Goldstone bosons of that spontaneously broken
chiral symmetry. The merit of standard $\chi$PT is first, that it is based on an effective Lagrangian density
constructed in accordance with all chiral constraints of QCD and second, that it permits a systematic
evaluation applying formal power counting rules \cite{book:Weinberg}. In $\chi$PT the finite but small values of
the up and down quark masses $m_{u,d} \simeq $ 10 MeV are taken into account as a small perturbation defining the
finite masses of the Goldstone bosons. The smallness of the current quark masses on the typical chiral scale
of $1$ GeV explains the success of standard $\chi$PT.

It is of course tempting to generalize the chiral $SU(2)$ scheme to the $SU(3)$ flavor group which
includes the strangeness sector, the main theme of this talk. To construct the appropriate
chiral $SU(3)$ Lagrangian is mathematically straightforward
and has been done long ago (see e.g. \cite{book:Weinberg}). The mass $m_s \simeq (10-20)\,m_{u,d}$ of the strange quark, though
much larger than the up and down quark masses, is still small on the typical chiral scale of
1 GeV . The required approximate Goldstone boson octet is readily found with the pions, kaons and the eta-meson.
Nevertheless, important predictions of standard $\chi $PT as applied to the $SU(3)$ flavor group are in stunning
conflict with empirical observations. Most spectacular is the failure of the Weinberg-Tomozawa theorem \cite{WT:1,WT:2}
which predicts an attractive $K^-$-proton scattering length, rather than the observed large and repulsive
value. Progress can be made upon accepting a crucial observation that the power counting rules
must not be applied to a certain subset of Feynman diagrams. Whereas for irreducible diagrams
the chiral power counting rules are well justified, this is no longer necessarily the case for the irreducible diagrams
\cite{Weinberg}. The latter diagrams are enhanced as compared to irreducible diagrams and
therefore require a systematic resummation scheme in particular in the strangeness sectors.
The $\chi$-BS(3) approach, which is based on the coupled-channel Bethe-Salpeter scattering equation,
was developed in great detail over the last years \cite{Hirschegg,Lutz:Kolomeitsev,LK03}. It constitutes a rigorous
effective field theory based on the above ideas. All results presented and discussed in this
talk are based on computations within the $\chi$-BS(3) approach.

Once the microscopic interaction of the Goldstone bosons with the constituents of nuclear
matter is understood one may study the properties of Goldstone bosons in nuclear matter.
Not only from an experimental  but also from a theoretical point of view the pions and kaons, the
lightest excitation of the QCD vacuum with masses of $140$ MeV and $495$ MeV respectively, are
outstanding probes for exciting many-body dynamics. The Goldstone bosons are of particular
interest since their in-medium properties reflect the structure of the nuclear many-body ground
state. For example at high baryon densities one expects the chiral symmetry to be restored. One
therefore anticipates that the Goldstone bosons change their properties substantially as
one compresses nuclear matter.

Even though in the $SU(3)$ limit of QCD with degenerate current quark
masses $m_u=m_d=m_s$ the pions and kaons have identical properties with respect to the strong
interactions, they provide very different means to explore the nuclear many-body system.
This is because the $SU(3)$ symmetry is explicitly broken  by a nuclear matter
state with strangeness density zero, a typical property of matter produced in the
laboratory. A pion, if inserted into isospin degenerate nuclear matter, probes rather
directly the spontaneously broken or possibly restored chiral $SU(2)$ symmetry.
A kaon, propagating in strangeness free nuclear matter, looses its Goldstone boson character
since the matter by itself explicitly breaks the $SU(3)$ symmetry. It is subject to three
different phenomena: the spontaneously broken chiral $SU(3)$ symmetry, the explicit symmetry
breaking of the small current quark masses and the explicit symmetry breaking of the
nuclear matter bulk. The various effects are illustrated by recalling the effective pion
and kaon masses in a dilute isospin symmetric nuclear matter gas. The low-density
theorem \cite{dover,njl-lutz} predicts mass changes $\Delta m_\Phi^2$ for any meson $\Phi $ in
terms of its isospin averaged s-wave meson-nucleon scattering length $a_{\Phi N}$
\begin{eqnarray}
\Delta m_\Phi^2 =-4\,\pi \left(1+\frac{m_\Phi}{m_N}\right) a_{\Phi N}\,\rho
+ {\mathcal O} \left( \rho^{4/3} \right)
\label{LDT}
\end{eqnarray}
where $\rho $ denotes the nuclear density.
According to the above arguments one expects that the
pion-nucleon scattering length $a_{\pi N} \propto m_\pi^2$ must vanish in the chiral $SU(2)$ limit since
isospin symmetric nuclear matter conserves the Goldstone boson character of the pions at least at
small densities. On the other hand, kaons loose their Goldstone boson properties and therefore one expects
$a_{K N} \propto m_K$ and in particular $a_{K^- N} \neq a_{K^+ N}$. This is demonstrated by the
Weinberg-Tomozawa theorem which predicts the s-wave scattering length in terms of the chiral order
parameter $f_\pi \simeq 93$ MeV :
\begin{eqnarray}
a_{\pi N} = 0 + {\mathcal O} \left( m_\pi^2\right) \,, \qquad
a_{K^\pm N} = \mp \frac{m_K}{4\pi \,f_\pi^2}+ {\mathcal O} \left( m_K^2\right) \,.
\label{WT-theorem}
\end{eqnarray}
In the pion sector the Weinberg-Tomozawa theorem (\ref{WT-theorem}) is beautifully confirmed by the
smallness of the empirical isospin averaged pion-nucleon scattering length $a_{\pi N} \simeq -0.01$ fm.
In the kaon sector the Weinberg-Tomozawa theorem misses the empirical scattering $K^+$ nucleon scattering length
$a_{K^+ N} \simeq - 0.3 $ fm by about a factor of three. Even more spectacular is the disagreement
of Weinberg-Tomozawa term in the $K^-$ case where (\ref{WT-theorem}) predicts $a_{K^- N} \simeq + 0.9 $ fm
while the empirical $K^-$ nucleon scattering length is about
$a_{K^- N} \simeq (- 0.6 +i\,1.1 ) $ fm.
Whereas  in conjunction with the low-density theorem the Weinberg-Tomozawa theorem predicts a decreased
effective $K^-$ mass, the empirical scattering length unambiguously states that there must be
repulsion in the $K^-$ channel at least at very small nuclear densities. The antikaon-nucleon scattering
process is complicated due to the open inelastic $\pi \Sigma$ and $\pi \Lambda $ channels. This is reflected
in the large imaginary part of the empirical $K^-$ nucleon scattering length. A quite rich variety of
phenomena arises from the presence of both the s-wave $\Lambda(1405)$ and p-wave $\Sigma(1385)$ resonances
just below, and the  d-wave $\Lambda(1520)$ resonance not too far above the antikaon-nucleon threshold.

In nuclear matter there exist multiple modes with quantum numbers of the $K^-$ resulting from the
coupling of the various hyperon states to a nucleon-hole state \cite{KVK95}. As a consequence the $K^-$ spectral
function shows a rather complex structure as a function of baryon density, kaon energy and momentum.
This is illustrated by recalling the low-density theorem as applied for the energy dependence of the kaon
self energy $\Pi_{\bar K}(\omega , \rho)$.  At zero antikaon momentum the latter,
\begin{eqnarray}
\Pi_{\bar K}(\omega , \rho) =
- 4\,\pi \left( 1+ \frac{\omega }{m_N}\right) \,f^{\rm s-wave}_{\bar K N}(m_N + \omega )\,\rho
+{\mathcal O} \left(  \rho^{4/3}\right) \,,
\label{LDT-energy}
\end{eqnarray}
is determined by the s-wave kaon-nucleon scattering amplitude
$f^{\rm s-wave}_{\bar K N}(\sqrt{s})$. A pole contribution to $\Pi_{\bar K}(\omega , \rho)$
from a hyperon state with mass $m_H$, if sufficiently strong, may lead to a $K^-$ like state of approximate energy
$ m_H-m_N$. Most important are the $\Lambda (1405)$ s-wave resonance and the $\Sigma(1385)$ p-wave
resonance. An attractive modification of the antikaon spectral function was already anticipated in the 70's by the many
K-matrix analyses of antikaon-nucleon scattering (see e.g. \cite{A.D.Martin}) which predicted
considerable attraction in the subthreshold s-wave $K^-$ nucleon scattering amplitudes. In
conjunction with the low-density theorem (\ref{LDT-energy}) this leads to an attractive antikaon spectral
function in nuclear matter. As was pointed out first in \cite{ml-sp} the realistic evaluation of the
antikaon self energy in nuclear matter requires a self-consistent scheme. In particular the feedback effect
of an attractive antikaon spectral function on the antikaon-nucleon scattering process was found to
be important for the $\Lambda(1405)$ resonance structure in nuclear matter.

\section{Antikaons and hyperon resonances at SIS energies}

\begin{figure}[t]
\begin{center}
\includegraphics[width=14cm,clip=true]{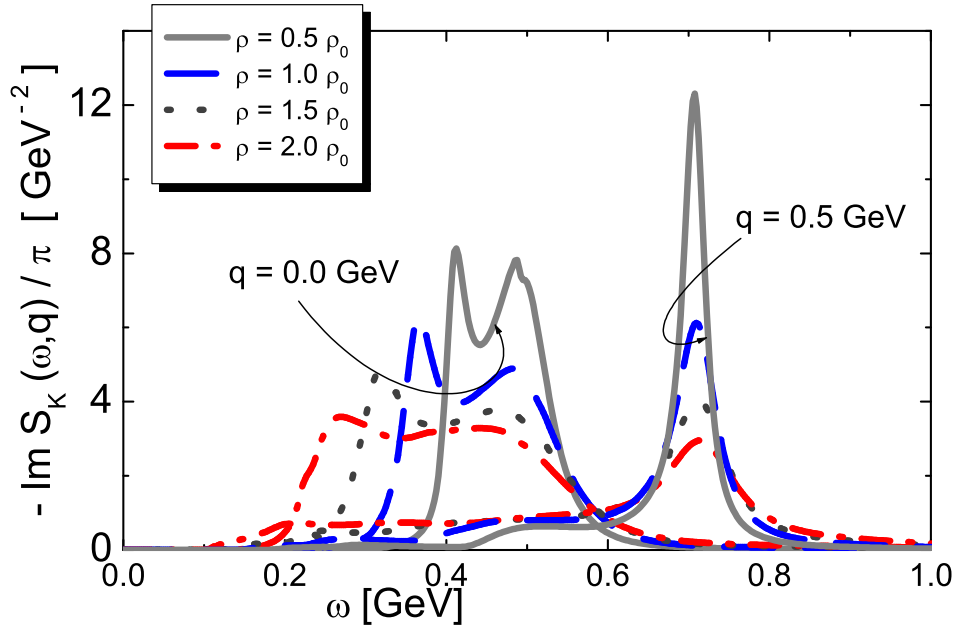}
\end{center}
\caption{Antikaon spectral function at momentum $q= 0$ MeV and $q=500$ MeV and
various nuclear matter densities. The results are based on the computation presented
in \cite{Lutz:Korpa}.}
\label{fig:kaon-spec-koch}
\end{figure}

The quantitative evaluation of the antikaon spectral function in nuclear matter is a
challenging problem. It should be based on a solid understanding of the antikaon-nucleon
scattering process in free space. Based on a description of the data set one obtains
a set of antikaon-nucleon scattering amplitudes. The present data set for antikaon-nucleon
scattering leaves much room for different theoretical extrapolations to subthreshold energies
\cite{A.D.Martin,martsakit,kim,sakit,gopal,oades,Juelich:2,dalitz,Kaiser,Ramos,keil}.
As a consequence the subthreshold $\bar K N$ scattering amplitudes of different analyses may
differ by as much as a factor of two \cite{Kaiser,Ramos} in the region of the $\Lambda (1405)$
resonance. Thus it is of crucial importance to apply effective field theory methods in order
to control the uncertainties. In particular constraints from crossing symmetry and chiral symmetry
should be taken into account. In  Fig. \ref{fig:kaon-spec-koch} the antikaon spectral function
at momenta $q= 0$ MeV and $q=500$ MeV evaluated at several nuclear densities is shown
\cite{Lutz:Korpa}. The results are based on antikaon-nucleon scattering amplitudes obtained
within the chiral coupled-channel effective field theory \cite{Lutz:Kolomeitsev}, where s-, p-
and d-wave contributions were considered. The many-body computation \cite{Lutz:Korpa} was
performed in a self-consistent manner respecting in addition constraints arising from
covariance. Fig. \ref{fig:kaon-spec-koch} illustrates that at zero momentum the spectral
function acquires a rather broad distribution as the nuclear density increases, with support
significantly below the free-space kaon mass. This reflects
the opening of many inelastic channels. In nuclear matter the antikaon couples strongly to
hyperon-resonance nucleon-hole states. Most important are the $\Lambda(1405)$ and $\Sigma(1385)$
resonances. At moderate momenta $q= 500$ MeV the spectral function looks significantly
different as compared to the one at $q= 0$ MeV. It is characterized by a peak close to
$\omega = \sqrt{m_K^2+q^2}$ and a pronounced low-energy tail. As the density increases the
strength in the peak diminishes shifting more and more strength into the low-energy tail.

Of interest are the implications of results for antikaon and
hyperon-resonance spectral distributions in cold nuclear matter
for subthreshold antikaon production in heavy-ion reaction at
SIS energies. Strictly speaking such results do not apply for the conditions
present inside a fireball created after the collision of two heavy nuclei. The
quantitative description of such a process
requires a transport theoretical description capable to model physics of finite systems
off equilibrium. The latter requires as input hadronic cross sections, or more precisely
hadronic transition rates, that are taken from experiment or theory. Calculations performed
in the idealized world of infinite but hot and dense nuclear matter are nevertheless a useful
tool to discuss and understand effects on a qualitative level. In certain channels
transition rates that are typically derived from free-space cross sections may be significantly
changed in a hot and dense nuclear environment.

One may argue that if the antikaon spectral function evaluated for
infinite hot nuclear matter shows considerable strength at energies smaller than the
free kaon mass, it is energetically easier to produce antikaons in a fireball than if the
effective antikaon mass distribution was not reduced. In this case one may expect to see some
type of enhancement of antikaon production rates. It is a challenge to verify the above
simple but physical interpretation of the observed kaon and antikaon yields in terms of
transport model simulations. In \cite{Foerster} results of the KaoS collaboration
for the kaon and antikaon transverse spectra for Au+Au at $E_{\rm beam}$=1.5 AGeV are shown.
The spectra shown are typical for most observed particle multiplicities. They are fitted well
by a simple Boltzmann distribution
\begin{eqnarray}
\sim \exp \big( - (E_{cm}-m_K)/T \big)\,,
\label{}
\end{eqnarray}
in terms of an effective temperature parameter $T$. The effective temperatures of the $K^+$
are about 20 MeV larger than the ones for the $K^-$ \cite{Foerster}. It is important to
realize that the Boltzmann type behavior of the spectra does not imply that the kaons and
antikaons are necessarily equilibrated. Estimates for the kaon mean free path, which are
based on the empirical kaon-nucleon cross section, suggest that the kaons do not have a
chance to reach equilibrium conditions in a typical heavy-ion collision at SIS energies.
That may be different for antikaons as will be discussed in some detail below.

Since at subthreshold energies the mean
energy available by one nucleon is not sufficient to produce an antikaon in an elementary
nucleon-nucleon collision by definition, one visualizes the production process to occur
in a series of successive reactions
\cite{Cassing:Mosel,Li:Brown,Aichelin:Hartnack,Aichelin:Oeschler:Hartnack},
\begin{eqnarray}
&& N\,N \to N \,\Delta \to N\,N \,\pi \,, \quad
N\,N \to N \,Y\,K \,,\quad
\pi \,Y \to \bar K \,N \,,
\label{prod-mech}
\end{eqnarray}
where $Y = \Lambda(1115), \Sigma (1185)$. Here we do not
intend to provide a complete listing of reactions included in transport model
simulations rather we would like to discuss the main effects qualitatively. A heavy-ion
reaction produces, besides photons and leptons, dominantly pions, the lightest
hadronic degrees of freedom available. For SIS energies at GSI
this production is thought to be driven by the isobar production process
$N\,N \to N \,\Delta \to N \,N \pi $. The kaon production is determined by the primary
$N\,N \to N \,Y\,K $ reaction far above threshold but by the secondary reaction
$\pi \,N \to K \,Y$ at subthreshold energies. Because of strangeness conservation the antikaon
production threshold in nucleon-nucleon collisions is much larger than the
threshold of kaon production. The former is determined by the $N\,N \to N\,N\,K\,\bar K$
reaction leading to
\begin{eqnarray}
\sqrt{s}^{(\bar K)}_{\rm thres}-\sqrt{s}^{(K)}_{\rm thres} = m_K +m_N-m_\Lambda \simeq
320 {\rm MeV} \,.
\end{eqnarray}
It is therefore plausible that antikaons are dominantly
produced in the secondary $\pi \,Y \to \bar K \,N$ reaction in particular at subthreshold
energies. Consequently these reactions received considerable attention in the study of
subthreshold antikaon production \cite{Cassing:Mosel,Li:Brown,Schaffner}. A clear hint that
the antikaon production in nucleus-nucleus collisions is indeed driven by the secondary
$\pi\, Y \to \bar K\, N$ reaction follows from the centrality dependence of the measured
ratio $K^-/K^+$  \cite{Foerster}. The number of nucleons, $A_{\rm part}$ that participate
in a Ni on Ni or Au on Au collision is a measure for the inverse size of the impact parameter.
Since both, the kaon and antikaon
yields depend strongly on the impact parameter \cite{Menzel:KaoS} the empirical observation
that the ratio of the yields is almost insensitive to it signals that the production
mechanisms of the kaons and antikaons must be strongly related.

So far most transport model simulations of antikaon production rely on rough approximations.
For instance the complicated antikaon spectral function is substituted by a quasi-particle
ansatz parameterized in terms of an effective antikaon mass that drops basically linearly
with the nuclear density \cite{Cassing:Mosel,Li:Brown}. Technically this is implemented by
incorporating an appropriate mean field modifying the trajectories of the
antikaon between collisions in the fireball. The width of the quasi particle is
then modelled by including the inelastic cross sections like $\bar K N \to \pi Y$. For cross
sections involving antikaons in initial or final states simple substitution rules like e.g.
$\sqrt{s} \to \sqrt{s} - \Delta m_{\bar K}$ are applied. Such computations confirm
the expectation that the antikaon yield is quite sensitive to an attractive mean field of
the antikaon \cite{Cassing:Mosel,Li:Brown}. Runs where an attractive mean field and the absorption
cross sections are both switched off or on are reported to be fairly close to the measured
multiplicities \cite{Cassing:Mosel,Li:Brown}. Neglecting, however the attractive mean field
but including the antikaon absorption cross sections underestimates the antikaon yield up to a
factor 5 \cite{Cassing:Mosel,Li:Brown}.

A more sophisticated simulation was performed by Schaffner, Effenberger and
Koch \cite{Schaffner} in which the consequences of two effects were studied.
First, the effect of in-medium modifications of the $\pi Y \to \bar K N$
reactions was studied. And second, a momentum dependence of the antikaon
potential was considered. The momentum dependence was chosen such that at
moderate antikaon momenta of about 300-400 MeV the attractive mean field was basically
switched off. This was motivated by microscopic many-body evaluations of the antikaon
self energy which predicted that type of behavior \cite{ml-sp}. The in-medium modification
of the production cross sections $\pi Y \to \bar K N$ were assumed to be dominated by
the Pauli blocking effect \cite{Koch} that shifts the $\Lambda(1405)$ resonance from
$\sqrt{s}= 1405$ MeV to larger energies, about $\sqrt{s}\simeq 1490$ MeV at nuclear
saturation density. As a consequence the authors report an enhancement factor
for the $\pi \Sigma \to \bar K N$ reaction of about 20 close to $\sqrt{s} =1490$ MeV.
Both effects, the attractive mean field and the increased
$\pi Y \to \bar K N$ cross section gave rise to a similar effects in the antikaon
yield. Together the two mechanism predict an enhancement factor of about 4 as compared
to a computation that applies the empirical cross sections and discards any mean-field
effects. However, as was pointed out by Schaffner, Effenberger and
Koch \cite{Schaffner}, once the momentum dependence of the attractive mean field was
incorporated together with the fact that in their model the enhancement of the
cross section disappears already at moderate temperatures $T \simeq 80$ MeV, the
results are quite close to the reference calculation
with no medium effects. All together the empirical antikaon yields are not described, they
are underestimated by about a factor 3-4.

In a more recent work by Aichelin, Hartnack and Oeschler
\cite{Aichelin:Hartnack,Aichelin:Oeschler:Hartnack} a
further interesting aspect was emphasized. Since the antikaon production is
driven by the $\pi Y \to \bar K N$ reaction the total antikaon yield should
be sensitive to the kaon production mechanism. This follows since the number of kaons and
hyperons available are comparable due to the typical production reaction (\ref{prod-mech}).
A repulsive mass shift for the kaon can therefore compensate for the effect implied by an
attractive mean field for the antikaon. We would argue, however, that the computation
by Aichelin and Hartnack \cite{Aichelin:Hartnack} most likely overemphasizes this effect due
to a too large mass shift for the kaon (about 70 MeV at 2 $\rho_0$). This objection
is based on a recent calculation based on the kaon-nucleon s- and p-wave phase shifts
\cite{Lutz:Kolomeitsev:Korpa} demonstrating that a kaon produced with
momenta of about 400 MeV at 2 $\rho_0$ the repulsive mass shift is only
about 45 MeV, much smaller than the value of 70 MeV used in \cite{Aichelin:Hartnack}.
The conclusions of Aichelin, Hartnack and Oeschler
\cite{Aichelin:Hartnack,Aichelin:Oeschler:Hartnack} depend also
sensitively on the magnitude of the attractive mean fields for the nucleons and hyperons
as well as on the poorly known $N\,\Delta \to N \,Y \,K$ reaction rates. In order to
reduce such uncertainties it would be useful to describe not only total yields of kaons,
antikaons and hyperons but at the same time achieve quantitative agreement with the
azimuthal emission pattern \cite{Li:Ko:Li,Fuchs:flow,Bratkovskaya}. Aichelin, Hartnack
and Oeschler \cite{Aichelin:Hartnack,Aichelin:Oeschler:Hartnack} affirm that the
$K^-/K^+$ ratio is rather insensitive to the details of the kaon production and therefore should
be considered as a superior observable to test the antikaon dynamics.
However, their claim, that including an attractive mean field for the antikaon into the simulation
does not affect the total antikaon yield, remains puzzling. This contradicts the typical
enhancement factors 2-4 as were obtained in previous simulations
\cite{Cassing:Mosel,Li:Brown,Cassing:review,Schaffner}.
In their work \cite{Aichelin:Oeschler:Hartnack} the possible importance of
in-medium modified cross sections is studied. Within their
scenario, which does not consider a momentum dependence of the antikaon mean field,
they observe an enhancement of the $K^-/K^+$ ratio by a factor 2 if the
$\pi Y \to \bar K N$ cross sections are enlarged uniformly by a factor three. However, the
impact parameter dependence of that ratio is affected strongly
by this ad-hoc procedure leading to a dependence that appears incompatible
with the behavior shown in \cite{Foerster}. It is argued that the
antikaons are produced at a late stage of the collision even though the difference
of absorption and production rates shows a clear maximum in the high density
($\rho > 1\,\rho_0$) initial phase.
Increasing the $\pi Y \to \bar K N$ cross sections only affects the antikaon absorption
process at the late stage of the fireball expansion since the typical pions and hyperons
do not have sufficient kinetic energy to produce antikaons anymore. This would lead to a drop
of the $K^-/K^+$ ratio with increasing impact parameter.

\begin{figure}[t]
\begin{center}
\includegraphics[width=14cm,clip=true]{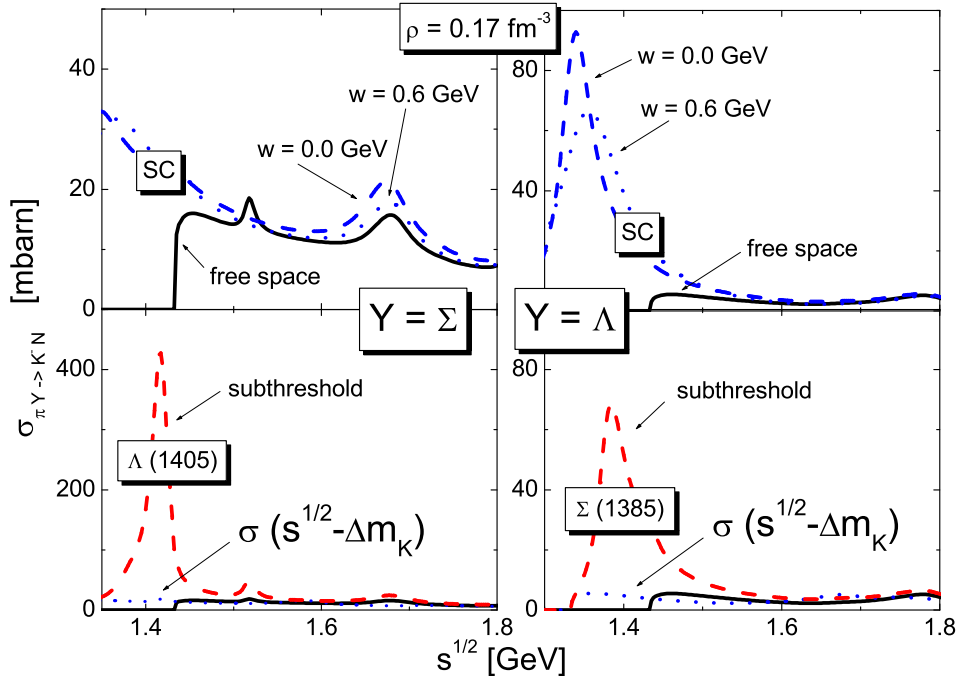}
\end{center}
\caption{The upper panels show the pion induced cross sections of
antikaons obtained in a self-consistent many-body evaluation (dashed for $\vec w=0$ MeV
and dotted lines for $\vec w=600$ MeV) at nuclear saturation density as
compared to the free-space cross sections (full lines). The lower panels give the results
of schematic evaluations. The dotted lines are the free-space cross section shifted in
$\sqrt{s}$ by $\Delta m_{\bar K}=-100$ MeV. The dashed lines follow with computations that
are based on the subthreshold free-space amplitudes as predicted by the $\chi$-BS(3) approach
together with a final state phase space evaluated with a reduced kaon mass.}
\label{fig:in-medium-cross}
\end{figure}

The above discussion implies that it is important to obtain an improved
understanding of the $\pi Y \to \bar K N$ cross sections as they may change in a
nuclear environment \cite{Schaffner}. To the extend that the latter reaction is of
crucial importance for the subthreshold antikaon production at SIS energies, chiral
symmetry plays a decisive role in the understanding of the antikaon yield in heavy-ion
reactions. In Fig. \ref{fig:in-medium-cross} the result of a computation based on the
self-consistent scheme introduced in \cite{Lutz:Korpa,LH02} are presented. In the upper panel
we show the isospin averaged $\pi \Sigma \to \bar K N$ and $\pi \Lambda \to \bar K N$
cross sections where the in-medium scattering amplitudes are used together with the
antikaon spectral function defining the
final-state phase space. The latter ones are taken from \cite{Lutz:Kolomeitsev,Lutz:Korpa}.
Comparing the solid with the dashed and dotted
lines one realizes a dramatic enhancement for the $\pi \Lambda$ reaction but a much
more moderate enhancement for the $\pi \Sigma $ reaction. The dashed and dotted lines
correspond to the situation where the sum of initial three-momenta is 0 MeV and 600 MeV
respectively. To explain the source of this effect the figure shows in its lower
panel the same cross sections evaluated in two different schematic ways. The dotted
lines give the result obtained according to the prescription
$\sqrt{s} \to \sqrt{s} - \Delta m_{\bar K}$ as commonly applied in transport
model simulations. It is used $\Delta m_{\bar K} =- 100$ MeV typically for the amount of
attraction found for the antikaon at saturation density. The cross sections remain as
small as they are in free-space. A striking enhancement is shown by the dashed lines in
the lower panel. Here we use the free-space scattering amplitude and evaluate the cross section
with the final state phase space determined by a reduced kaon mass of 394 MeV. As a result
this cross section probes the scattering amplitudes at subthreshold energies a kinematical
region where they are not directly constrained by the scattering data. Since the $\chi-$BS(3)
scheme predicts considerable strength in the subthreshold amplitudes from the s-wave
$\Lambda(1405)$ and p-wave $\Sigma (1385)$ resonances the former cross sections are
dramatically enhanced as compared to the free-space cross sections
Comparing the lower
with the upper panel demonstrates that the latter cross section, though providing a
simple physical interpretation of the enhancement, do not adequately reproduce the
full computation as it arises in a self-consistent framework. In particular the large
cross section of the $\pi \Sigma \to \bar K N$ reaction predicted by the free-space amplitudes
is significantly reduced in the self-consistent scheme. Here one should note that
the s-wave and p-wave final-state phase space factors probe the antikaon spectral function
in different ways. Thus, the net result is a combined effect depending on the in-medium
amplitude and a projection of the antikaon spectral function that depends on the angular
momentum.

The moderate enhancement for the $\pi \Sigma \to \bar K N$ reaction confirms the results of our
previous work \cite{ml-sp}, in which it was pointed out that a self-consistent
antikaon dynamics does not generate the large enhancement factor predicted by a scheme that
considers Pauli blocking only \cite{Koch,Schaffner}. Since an enhancement factor that is driven
by the Pauli blocking effect will eventually disappear once a finite temperature is allowed
for, it can not explain enlarged cross sections to be used in heavy-ion reactions in any case.
The striking effect induced by the $\Sigma (1385)$ resonance in the $\pi \Lambda \to \bar K N$
reaction is novel. It is the result of our detailed analysis of the kaon and antikaon
scattering data which predicts that there is a strong coupling of the $\Sigma (1385)$
resonance to the $\pi \Lambda$ and $\bar K N$ channels at subthreshold energies. The
self-consistent many-body computation presented here suggests that this strong coupling
persists in the nuclear medium giving rise to the large enhancement factor found for the
in-medium $\pi \Lambda \to \bar K N$ reaction. This effect should have important consequences
for the antikaon yield in nucleus-nucleus collisions at SIS energies.

As demonstrated in
\cite{Aichelin:Oeschler:Hartnack} the ad-hoc increase of the $\pi Y \to \bar K N$
cross section away from its free-space limit does increase the total antikaon yield.
This demonstrates that transport model simulations of the antikaon yield that are based
on free-space cross sections do not reach equilibrium conditions. Of course beyond
some critical enhancement factor one would expect the yield to saturate implying that
phase space is populated statistically. To put it in another way an enhancement factor
40 will not increase the yield by a factor of 40. However, we argue that a strong
enhancement factor is rather welcome since that implies that the enhancement should
be strong enough exceeding the critical value required for driving the antikaons towards
equilibrium conditions even at smaller density or equivalently larger impact parameter.
Therefore we would expect that the impact parameter dependence of the $K^-/K^+$ yield
should be weak as seen \cite{Foerster} and predicted by
the statistical model of Cleymans, Oeschler and Redlich \cite{Cleymans:Oeschler:Redlich}.
Here we would deviate from the line of arguments put forward by Aichelin, Hartnack
and Oeschler who argued that with their ad-hoc enhancement the impact parameter
dependence of the $K^-/K^+$ yield decreases with increasing impact parameter. Since
the unscaled cross sections predict a flat behavior in their simulation, more consistent
with \cite{Foerster}, one may favor the free-space cross sections.
This conclusion would be based, however, on a simple energy independent enhancement factor
of the $\pi \,Y \to \bar K \,N$ transition rate. In contrast Fig. \ref{fig:in-medium-cross}
suggests that even at a rather late stage of the fireball expansion the pions and hyperons need
very little kinetic energy to produce a virtual antikaon. That should lead to a flattening
of the number of antikaons with increasing reaction time contradicting the results shown
by Aichelin, Hartnack and Oeschler \cite{Aichelin:Oeschler:Hartnack}.

It would be very useful to incorporate the dynamics of the $\Sigma (1385)$ into
transport model simulations. For a first attempt see \cite{Laura}.
If it can be confirmed that that the description of
the antikaon yield requires an in-medium enhancement of the
$\pi\, Y \to \bar K \, N$ rates we would interpret this as a clear signal for an attractive
shift in the antikaon mass distribution. Only if the in-medium $\bar K N$ phase space has
significant overlap with the $\Sigma(1385)$ resonance the $\pi\, \Lambda \to \bar K \, N$
rate can be enhanced significantly.

\section{Conclusion}

We reviewed the relevance of the chiral SU(3) symmetry for subthreshold
antikaon production in heavy-ion collisions \cite{Senger}. It is argued that
chiral symmetry plays a decisive role since it helps to control the complicated
interaction of antikaons with the constituents of matter. Results, that are based
on the chiral coupled-channel effective field theory, have interesting consequences for antikaon
propagation in dense nuclear matter. For the antikaon spectral function a pronounced dependence
on the three-momentum of the antikaon is predicted. The spectral function shows typically a rather
wide structure invalidating a simple quasi-particle description. For instance at
$\rho = 0.17$ fm$^{-3}$ the strength starts at the quite small energy, $\omega \simeq $ 200 MeV.
Furthermore, at nuclear saturation density attractive mass shifts for the
$\Lambda(1405)$, $\Sigma (1385)$ and $\Lambda(1520)$ are foreseen. The hyperon states are
found to show at the same time moderately increased decay widths.

The importance of the p-wave $\Sigma (1385)$ resonance for the antikaon production in heavy
ion collisions as studied at GSI was pointed out. The presence of the latter resonance
implies a substantially enhanced $\pi \Lambda \to \bar K N $ reaction rate in nuclear matter.
Contrary to naive expectations only a small enhancement of the $\pi \Sigma \to \bar K N $
reaction was derived even though it couples to the s-wave $\Lambda (1405)$ resonance. It was
emphasized that the incorporation of the $\Sigma (1385)$ resonance into transport model
simulations of nucleus-nucleus collisions at SIS energies should help to obtain a quantitative
and microscopic understanding of the empirical antikaon yields. To help understanding this
dynamics it is desirable to measure the $\Sigma(1385)$ yield at SIS energies.

\end{document}